# Comment on: "Dynamics of Phospholipid Membranes beyond Thermal Undulations"

## Rony Granek[*]

*The Stella and Avram Goren-Goldstein Department of Biotechnology Engineering,
and the Ilse Katz Institute for Meso and Nanoscale Science and Technology,
Ben-Gurion University of The Negev, Beer Sheva 84105, Israel*

**ABSTRACT:** I introduce an alternative interpretation for the short-time anomalous diffusion observed by Gupta *et al.* using neutron spin echo. The interpretation involves the Brochard-Lennon "red blood cell" mode to describe membrane thickness fluctuations, and yields an anomalous diffusion exponent 1/3 that is consistent with the experimental results.

The result of Gupta *et al.*[1] of a distinct exponent, $\nu \approx 0.26 \pm 0.03$, governing the (very short time) stretched exponential relaxation of the intermediate scattering function, and the transverse anomalous mean square displacement (MSD) of a membrane segment, is both intriguing and fascinating. It has been qualitatively interpreted by the authors[1] as a molecular trapping-like effect that leads to powerlaw tails of escape waiting time distribution, as described by the continuous time random walk (CTRW) formalism. Their explanation, based on molecular dynamics simulations[2], is associated with the two-dimensional (2D) motion of lipids that is linked to local membrane density fluctuations. Here I would like to offer an alternative explanation.

There have recently been suggestions that membrane thickness fluctuations are detectable by neutron spin echo (NSE). Nagao and co-workers[3] used a phenomenological expression for the relaxation rate, based on an elastic solid-like response to thickness deformations, which is added to the conventional Zilman-Granek (ZG) expression[4]. Here I assume that for thickness deformations of wavelengths shorter than, or comparable to, the membrane half-thickness, a liquid-like response of the phospholipid tails is possible, within a mode similar to the Brochard-Lennon (BL) "red blood cell" mode[5]. This mode corresponds to a peristaltic-like deformation of the membrane thickness with a distinct wavenumber $k$, and its relaxation rate $\omega$ scales as $\sim k^6$. The latter is a consequence of the bending energy of each of the two leaflets, which contributes $\sim \kappa_m k^4$ where $\kappa_m$ is the bending modulus of a monolayer leaflet, and the material-conservative hydrodynamic flow of the hydrocarbon tails that contributes a factor of $\sim k^2$. The resulting relaxation rate[5] takes the form $\omega = \left(\frac{\kappa_m d^3}{24\,\eta_{\text{eff}}}\right) k^6$, where $d$ is the bilayer thickness, $\eta_{\text{eff}}$ is the effective viscosity of the hydrocarbon tails, and the solvent viscosity has been neglected. Since obviously the hydrocarbon tails, being chemically bonded to their head-groups, are not truly free to flow independently of the head-groups, this effective hydrodynamic picture can only be relevant to peristaltic-like deformations of wavelengths shorter than the bilayer half-thickness. For such short wavelengths, by shearing the tails, they can be squeezed away from the narrow (bottleneck-like) regions of the deformation to fill space in the dilated regions. Notably, using continuous hydrodynamic approach is always a crude approximation at these molecular length-scales, nevertheless it has been proved useful time and again.

With this in mind, it is easy to deduce the time-dependent MSD of the thickness. Consider first the scaling hypothesis

$$\langle \delta d(t)^2 \rangle \equiv \langle (d(t) - d(0))^2 \rangle = \frac{k_B T}{\kappa_m} L^2 G(t/\tau_L) \qquad (1)$$

where $L$ is a membrane patch linear size, and $\tau_L = \omega^{-1}\left(k = \frac{\pi}{L}\right) \approx \frac{\eta_{\text{eff}}}{\kappa_m d^3} L^6$ is the longest bending relaxation time of this patch. Demanding that, at times $t \ll \tau_L$, $\langle \delta d(t)^2 \rangle$ become independent of $L$, implies $G(x) \sim x^{1/3}$ for $x \ll 1$, and so

$$\langle \delta d(t)^2 \rangle \simeq C \frac{k_B T \, d}{\kappa_m^{2/3} \eta_{\text{eff}}^{1/3}} \, t^{1/3}, \qquad (2)$$

where $C$ is a numerical prefactor, consistent with a power-spectrum $\sim \omega^{-4/3}$ predicted by BL[5]. A more complete calculation involving an integral expression allows to evaluate the numerical prefactor, $C \simeq 0.149$. As a consequence, the dynamic structure factor at large wavenumbers $q$ decays as a stretched exponential with an exponent 1/3, $S(q,t) \sim e^{-(\Gamma_q t)^{1/3}}$, and a decay rate (arising from the single leaflet relative MSD, $\langle \delta u(t)^2 \rangle = \langle \delta d(t)^2 \rangle / 2$)

$$\Gamma_q \simeq 5.2 \times 10^{-5} \frac{(k_B T)^3 d^3}{\kappa_m^2 \eta_{\text{eff}}} q^6 \, . \qquad (3)$$

The success to predict the anomalous diffusion (and stretching) exponent $\nu = 1/3$ is not sufficient to confirm this alternative explanation. The data of Gupta et al.[1] for DOPC, for example, corresponds to an effective viscosity that is roughly $\eta_{\text{eff}} \sim 1$ mPa s, which is a relatively low viscosity but perhaps still reasonable given the data and parameter uncertainties. This effective viscosity has been deduced by taking, in Eq. (2), $\kappa_m \simeq 10 \, k_B T$, estimated as half of the long wavelength bilayer bending modulus, $d \simeq 3.6$ nm, $t \simeq 3$ ns, and $\langle \delta d(t)^2 \rangle \simeq 27$ A$^2$; using $\kappa_m \simeq 5 \, k_B T$, for instance, yields a larger value, $\eta_{\text{eff}} \simeq 4$ mPa s. (The authors report on the 3D MSD $\langle \delta \vec{r}(t)^2 \rangle$, presumably deduced from $S(q,t)$ following the relation $S(q,t) = e^{-q^2 \langle \delta \vec{r}(t)^2 \rangle / 6}$ for the case of independent particles motion, and thus I have used the relation $\langle \delta \vec{r}(t)^2 \rangle = 3\langle \delta u(t)^2 \rangle = 3\langle \delta d(t)^2 \rangle / 2$ with $\langle \delta \vec{r}(t)^2 \rangle \simeq 40$A$^2$ for $t \simeq 3$ ns from the supporting information of Gupta et al.[1], Fig. S2.) The dependence of the MSD on the phospholipid identity could perhaps be checked given knowledge of $d$ and $\kappa_m$ for each lipid, although this comparison might be hampered if $\eta_{\text{eff}}$ is also sensitive to the lipid identity.

Despite the apparent consistency, the above interpretation implicitly assumes Gaussian statistics, thus contradicting the non-Gaussian behavior of $\delta d(t)$ found by Gupta et al. during the early time regime where the small exponent $\nu \approx 0.26 \pm 0.03$ is measured. (The Gaussian statistics is recovered only during the time regime where the exponent $\nu = 0.66$ is observed, consistent with the ZG model.) This fundamental contradiction calls for further investigation of the non-Gaussian nature of the fluctuations. Could it be a consequence of the NSE technique, associated with these extremely short times, and not an inherent property of the system? Or, is there another process – associated with a CTRW powerlaw tail – that is similar to the suggestion of the authors but one that involves bilayer deformations and not 2D lipid diffusion? Swelling the bilayer with a suitable oil could be a useful way to distinguish between the different mechanisms due to the sensitivity of Eqs. (2) and (3) to $d$.

**AUTHOR INFORMATION**

*E-mail: rgranek@bgu.ac.il

**REFERENCES**

1. Gupta, S.; De Mel, J. U.; Perera, R. M.; Zolnierczuk, P.; Bleuel, M.; Faraone, A.; Schneider, G. J., Dynamics of Phospholipid Membranes beyond Thermal Undulations. *The Journal of Physical Chemistry Letters* **2018,** *9* (11), 2956-2960.
2. Akimoto, T.; Yamamoto, E.; Yasuoka, K.; Hirano, Y.; Yasui, M., Non-Gaussian fluctuations resulting from power-law trapping in a lipid bilayer. *Physical review letters* **2011,** *107* (17), 178103.
3. (a) Ashkar, R.; Nagao, M.; Butler, P. D.; Woodka, A. C.; Sen, M. K.; Koga, T., Tuning membrane thickness fluctuations in model lipid bilayers. *Biophysical journal* **2015,** *109* (1), 106-112; (b) Woodka, A. C.; Butler, P. D.; Porcar, L.; Farago, B.; Nagao, M., Lipid Bilayers and Membrane Dynamics: Insight into Thickness Fluctuations. *Physical Review Letters* **2012,** *109* (5), 058102.
4. (a) Zilman, A.; Granek, R., Undulations and dynamic structure factor of membranes. *Physical review letters* **1996,** *77* (23), 4788; (b) Zilman, A. G.; Granek, R., Membrane dynamics and structure factor. *Chemical Physics* **2002,** *284* (1-2), 195-204.
5. Brochard, F.; Lennon, J., Frequency spectrum of the flicker phenomenon in erythrocytes. *Journal de Physique* **1975,** *36* (11), 1035-1047.